# A Rapid Scoping Review and Conceptual Analysis of the Educational Metaverse in the Global South: Socio-Technical Perspectives

Anmol Srivastava,

Creative Interfaces Lab, Dept. of Human-Centered Design, IIIT Delhi, anmol@iiitd.ac.in

This paper presents a conceptual insight into the Design of the Metaverse to facilitate educational transformation in selected developing nations within the Global South regions, e.g., India. These regions are often afflicted with socio-economic challenges but rich in cultural diversity. By utilizing a socio-technical design approach, this study explores the specific needs and opportunities presented by these diverse settings. A rapid scoping review of the scant existing literature is conducted to provide fundamental insights. A novel design methodology was formulated that utilized ChatGPT for ideation, brainstorming, and literature survey query generation. This paper aims not only to shed light on the educational possibilities enabled by the Metaverse but also to highlight design considerations unique to the Global South.

CCS CONCEPTS • Human-centered computing • Human computer interaction (HCI) • HCI theory, concepts and models

**Additional Keywords and Phrases:** Metaverse, Education, Global South, Socio-Technical Design

## 1 INTRODUCTION

This article conceptualizes the Design of Metaverse for the digital transformation of engineering education in the Global South (GS). A socio-technical system design (STS-D) approach has been adopted. STS-D blends the complex social and technical understanding for optimal performance and successful digital transformation of organizations [1]. The Metaverse is potentially the next phase of the internet and also for the digital transformation of numerous organizations [2], [3]. By amalgamating technologies, it envisions better prospects, economy, and immersive user experiences. It can transform the way people work, learn, socialize, and interact with each other, thus giving rise to a new socio-cultural phenomenon. While the Metaverse is envisioned to be the next digital transformation milestone, significant concerns and questions must be addressed, especially for the GS.

The GS refers to economically developing nations. The term includes Asia, Africa, Latin America, and Oceania [4]. One key aspect seen as crucial to the development of the GS is education and technology, which can lead to substantial growth in talent and modernization. Engineering education, in particular, plays an integral role in the economic growth of these nations and driving the knowledge economy. Reports [5] [6] have shown compelling evidence of how engineering skills correlate with economic advancement and development of nations and are indispensable in attracting investments and industrial flourishing. Engineering graduates, thus, are an essential resource to drive a country's knowledge-based capital and development [7].

In recent years, there has been a significant increase in engineering institutes in the GS [8]. However, many nations of the GS need to improve the quality of engineering education, which adversely affects their economy, development, and global competitiveness. Commonly cited problems are (a) poor quality of teaching and learning, (b) management and administration issues, (c) lack of soft-skill and internationalization amongst the graduates, and (d) poor infrastructure to support teaching and research activities [9], [10]. For example, around 1.5 million engineering students in India graduate every year from about 7000 engineering and technical institutes [11]. Many institutions need the proper infrastructure to support quality education and more teachers to transfer industry-ready skills to students [12]. Similar concerns have been reported in Africa, where insufficient output from training institutions, poor quality education, and lack of practical experience among graduates have been reported as the leading causes of lack of engineering capacity [13]. China, too, has reported concerns over the quality of engineering education [14].

Metaverse technologies [3], such as XR = Augmented/ Virtual/ Mixed Reality[15], decentralized networks [3], and blockchain [16], can act as enablers of the digital transformation of engineering education in the GS. However, there lies a critical paradox: Many nations of the GS struggle with socio-economic concerns and lack of a robust digital infrastructure [4], [17]. Some challenging aspects are the digital divide [18]–[20], digital literacy, and affordability, which must be considered thoroughly. In addition, there are aspects of cultural norms, language, and cultural representations [21]. These aspects pose significant barriers (and opportunities) for implementing the Metaverse in these nations. Thus, the moot questions are - how can the Metaverse truly aid engineering education in the GS? What are the design considerations for such educational Metaverse experiences?

To answer these questions, published literature was surveyed and reviewed. The inferences have been reported in this article and can be broadly considered while designing the educational Metaverse experience for engineering education in the GS. The paper aims to recognize both the revolutionary potential of the Metaverse and the genuine challenges that must be overcome in the GS context. In doing so, it seeks to contribute to a comprehensive understanding of how these emerging technologies can be responsibly and effectively integrated into the educational landscapes of developing regions.

## 2 METHODOLOGY

A Rapid scoping studies framework [22] was adopted primarily because of the need for published literature identified from an initial survey on socio-technical issues of educational Metaverse in the GS context. This framework provides a broad overview of the field by examining how the research is conducted, analyzing knowledge gaps, and identifying concept-related factors [23]. ChatGPT-4 [24] was used for brainstorming, ideation, prompt generation, and query generation for the literature survey. The use of Large Language Models (LLM) like ChatGPT in academics is a recent phenomenon under development with several ethical concerns [25]. However, relevant literature that provides suitable guidelines for utilizing these LLMs for conducting literature surveys [26], prompt generation [27], and design process [28], [29] were referred to formulate a suitable methodology for this study. Figure 1. presents a design research methodology developed and utilized in this study. Through an analysis of current theories, solutions, and gaps in the literature, this review lays the groundwork for an iterative process of Design, implementation, reflection, and refinement aimed at creating compelling, contextually appropriate Metaverse experiences. The study also lends insights from socio-technical systems design (STS-D) [1], information sciences [30], and Design [17] to arrive at a conceptual understanding.

The preliminary research questions (PRQ) were: **PRQ1** *How do we define the Metaverse?* **PRQ2** *What are the technical and social concerns of the Metaverse?* **PRQ3** *How can the Metaverse be utilized in education and engineering education?* Based on investigations of PRQs, research questions (RQ) were formulated to identify socio-technical aspects of the Metaverse and understand its implications on engineering education in the GS. The RQs are as follows:

**RQ1** *What are the socio-technical issues about the Metaverse for education in the GS?*
**RQ2** *What are the challenges for GS in implementing the Metaverse for engineering education?*

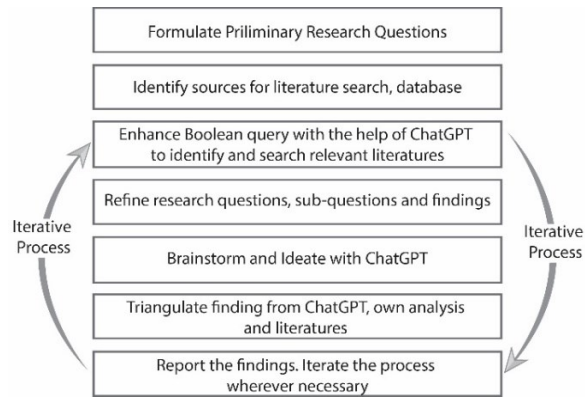

Figure 1: Design Research Methodology formulated for this study. An iterative process utilizing LLM like ChatGPT for brainstorming and query generation for literature review was adopted.

## 2.1 Rapid Scoping Review Process

Articles were selected based on the Title, Abstract, and Full-text. The inclusion criteria/ exclusion criteria for all surveys were the same. *Inclusion Criteria:* (a) Records published between 2019 and 2023; (b) Articles written in English; (c) Full texts. In some instances, grey literature was included if they discussed GS and the Metaverse. This was primarily done due to scant published academic work in this area. *Exclusion Criteria:* (a) Records published before 2019; (b) Articles not written in English.

A survey was first done on the Metaverse to investigate PRQ1 and PRQ2. The search was performed with the keyword "Metaverse" on Google Scholar. Relevant articles were shortlisted after reading the abstract and then read fully to gain a perspective of the Metaverse, its features, and its challenges. To answer *PRQ3*, i.e., the Metaverse for education, a search was performed on Google Scholar with the query ("Literature Review") AND ("metaverse") AND ("education" OR "learning"); ("Metaverse") AND ("education" OR "learning"). Table 1 depicts the search results on sources like Google Scholar, Science Direct, IEEE Xplore, and ACM DL. These initial surveys show that sufficient work exists on Metaverse for education and engineering education. It is also noted that there has been a rising interest in the Metaverse in these five years, primarily from an XR perspective that reports the use of AR, VR, and MR in education. However, the results that mainly reported XR technologies in education were excluded from this survey. This was done to limit the scope of the survey since there are numerous studies on XR.

Further, XR only forms a small subset of the Metaverse [3] that aims to amalgamate other AI and HCI technologies. Before 2019, it was noticed that the Metaverse mainly was referred to as virtual worlds. The emphasis was to understand how these 3D environments could influence education. The query was further refined to ("metaverse") AND ("engineering education"). For ***RO1*** and ***RO2***, ChatGPT 4.0 was used for developing a Boolean query based on the guidelines provided by the literature [26]. However, it was observed that these queries from ChatGPT-4 did not yield desirable search results; hence, the questions were modified to ("developing nations" OR "emerging economies") AND ("Metaverse") AND ("education"). While there are studies [31], [32] that discuss "developing nations," the term "GS" with the Metaverse was rarely used. The query ("socio-technical") AND ("Metaverse") yielded only a few literature that specifically mentioned socio-technical aspects of the Metaverse [33]–[35]. Table 2 presents relevant literature that addresses or discusses the RQs in parts.

Table 1: Search Query and Search Results

| S.No | Keywords Used | Source | Results | Year |
|---|---|---|---|---|
| 1 | ("Literature Review") AND ("metaverse") AND ("education" OR "learning") | Google Scholar | 4290 | 2019-2023 |
| 2 | ("metaverse") AND ("education" OR "learning") | Google Scholar | 13000 | 2019-2023 |
| 3 | ("metaverse") AND ("education" OR "learning") | ACM DL | 288 | 2019-2023 |
| 4 | ("metaverse") AND ("education" OR "learning") | IEEE Xplore | 236 | 2019-2023 |
| 5 | ("metaverse") AND ("education" OR "learning") | Science Direct | 499 | 2019-2023 |
| 6 | ("metaverse") AND ("engineering education") | Google Scholar | 619 | 2019-2023 |
| 7 | ("metaverse") AND ("engineering education") | ACM DL | Nil | 2019-2023 |
| 8 | ("metaverse") AND ("engineering education") | Science Direct | Nil | 2019-2023 |
| 9 | ("metaverse") AND ("engineering education") | IEEE Xplore | 8 | 2019-2023 |
| 10 | ("developing countries" OR "emerging economies") AND ("Metaverse") | Google Scholar | 1130 | 2019-2023 |
| 11 | ("Socio-technical") AND ("Metaverse") | Google Scholar | 446 | 2019-2023 |
| 12 | ("Socio-technical") AND ("Metaverse") | Science Direct | 21 | 2019-2023 |
| 13 | ("Socio-technical") AND ("Metaverse") | ACM DL | Nil | 2019-2023 |

Table 2: Relevant literature that focuses on the Global South Context of the Metaverse

| S.No. | Research Questions | Reference |
|---|---|---|
| RQ1 | *What are the socio-technical issues about the Metaverse for education in the GS?* | **Did not find any paper that explicitly addresses RQ1**. However, some relevant research work identified are: [35]–[37] |
| RQ2 | *What are the challenges for GS in implementing the Metaverse for engineering education?* | [31], [38] |

## 3 INSIGHTS FROM THE LITERATURE SURVEY

Common themes were identified from the literature survey of Metaverse for education and engineering education. Figure 2 presents a word cloud depicting recent literature's most used topics or keywords. Some commonly explored identified from the survey are presence and immersive learning [39], [40], visualization capabilities [39]–[45], remote collaboration, collaborative work [46]–[48], and accessibility [3] as main advantages of the Metaverse. Digital twins were often utilized for engineering education or in the context of higher education. From the GS perspective, specific papers [2], [31], [43], [48], [49] delved into design considerations for understanding cultural issues of the Metaverse, emphasizing that people from diverse backgrounds should be able to interact with each other via digital avatars. However, gaps exist in understanding how these avatars will be designed for developing nations and how to tackle cross-cultural issues that may arise in the Metaverse. There are concerns regarding access to digital infrastructure, the rural-urban divide, and the digital divide [31], [50], [51]. While the Metaverse can be utilized in urban settings, how it can reach the rural level is mainly reported to be a national policy matter. Interestingly, despite the general recognition of the Metaverse's potential in education, only [32], [35]–[37] focused explicitly on engineering education within the GS framework. From the survey, it is evident that while the papers present research agenda and design considerations for educational metaverse [2], [3], [33], [41], there is a lack of understanding as to how Metaverse can impact a broader socio-technical level from a GS perspective.

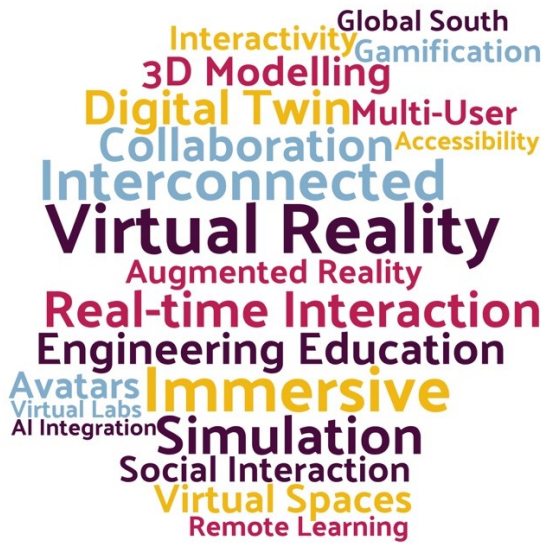

Figure 2: Most used topics or keywords mentioned in the literature survey.

Table 3 depicts the most common themes identified from the literature and their frequency of discussion. It can be noticed that themes like Accessibility and Digital Divide, Environmental Impact, Cross-cultural and Gender Issues aspects need more insights. Several sub-questions warrant deeper inquiry for Metaverse-based digital transformation in higher education, such as (a) What are the implications of the Metaverse on institutional/ organizational quality (efficiency, innovation, sustainability)? (b) How does it influence students' and employees' study quality/ working life (faculty or staff)? (c) How can it offer different forms of collaboration among stakeholders of higher engineering education in the GS? These questions are primarily derived from a STS-D [1] perspective.

Table 3: Common themes identified from the literature and their frequency of discussion

| S.No | Themes | Frequency of Occurrence (Estimate) |
|---|---|---|
| 1 | Virtual Reality (VR), Augmented Reality (AR), Mixed Reality (MR) technologies | High |
| 2 | Digital Ownership and Non-Fungible Tokens (NFTs) | High |
| 3 | Decentralization and Blockchain Technology | High |
| 4 | Privacy and Security Concerns | High |
| 5 | Virtual Economies and Cryptocurrencies | High |
| 6 | Spatial Computing and User Interfaces | Medium |
| 7 | Digital Identity and Avatars | Medium |
| 8 | Social Interaction and Community Building | Medium |
| 9 | Integration of Physical and Virtual Worlds | Medium |
| 10 | Legal and Ethical Implications | Medium |
| 11 | Accessibility and Digital Divide | Low |
| 12 | Environmental Impact of Technologies | Low |
| 13 | Cross-cultural/ Intergenerational interfaces | Low |
| 14 | Gender Issues, Class, etc. | Low |

## 4 DISCUSSION

The survey presents exciting insights regarding the educational affordances of the Metaverse and future research directions. It reveals concerns regarding cultural considerations that have been suggested. In addition, the published literature primarily reports issues such as digital infrastructure and digital divide while considering developing regions. The survey also reveals that more published literature with a specific focus on GS needs to be published. The existing studies often need a Design thinking lens to approach the Metaverse from a GS perspective, with more emphasis on technical considerations and, to a certain extent, cultural issues. Thus, thinking about designing a Metaverse from a human-centered design approach can be vital. This article presents a conceptual understanding of the Metaverse for GS utilizing a Design lens based on these insights. The discussions are shown in the following sub-sections.

### 4.1 Understanding the Design of Metaverse for the Global South

While the Metaverse is viewed as a potential disruption that can transform various sectors, economies, jobs, and education [3], there is a solid need to view it from the GS's perspective. As such, the whole notion of the Metaverse and its current drive, growth, and applications have been derived primarily from the developing nations or the Global North. The conditions, Design, and context required for shaping the Metaverse from the perspective of developing nations need to be included. An article [17] on Design for/by GS presents an insightful discussion on the GS, highlighting these developing nations' principles, practices, and needs. Designers must develop multi-pronged approaches rooted in the local context and needs when designing for the GS. This involves a collaborative, participatory design process involving local communities in decision-making. It also entails considering local people and experts and their experiences. Emphasis will be required on making a sustainable, equitable, socially acceptable, economically viable, multi-cultural, multi-lingual, and usable Metaverse. It will require materials and technology that are local, readily available, and feasible to implement. While the cost factor is always raised in discussions about the GS, it is crucial to understand these nations' diverse socio-economic and digital infrastructure. A broad perspective on the digital divide [18], [19], [52], [53] is required to understand that it is not just an infrastructural challenge but rather a complex social and cultural aspect. For example, it does not matter if the user has an internet connection if they cannot afford it, or uses a browser or connected devices, or if the language is understandable by them. It is also essential to avoid a 'one-size-fits-all all design' where the solutions of the Global North are imposed on the GS without sufficient understanding. This means that designers should consider local resources, skills, cultural norms, and contextual factors of these developing nations with "a non-western way of thinking and a sharper sensitivity towards gender, race, class, etc." [54].

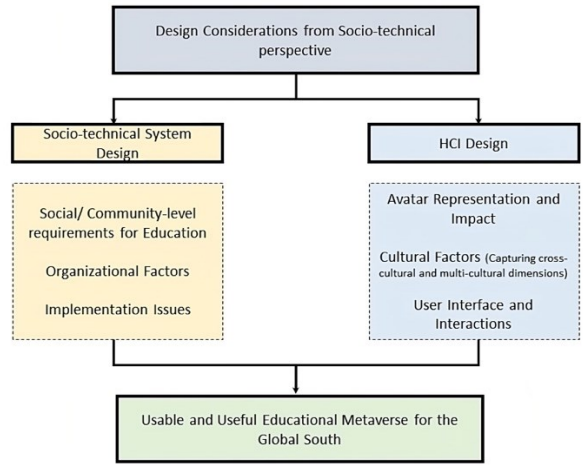

Figure 3: Design considerations for educational Metaverse in the Global South

Figure 3 depicts two primary considerations from a socio-technical Design and HCI design perspective. Both are important. The socio-technical design lens provides a broader understanding of the influence of technology on social and community levels. It entails considering factors that are often ignored when only technology is considered. This means that focusing alone on technology lowers productivity [1]. For the GS, it becomes an essential indicator not simply to jump on the Metaverse bandwagon but to closely understand how it will affect the organizations, people, job roles, students, faculty, skill development, and quality of education. Necessary measures will be required to implement this next digital transformation successfully. The HCI design perspective is crucial as it delivers a personalized experience to the users of the Metaverse. It can ensure a usable, practical, and user-friendly user experience that is more aligned with the needs of the GS.

## 4.2  Socio-technical and HCI Design Considerations

Several challenges exist from a socio-technical perspective for the Metaverse to be implemented in engineering education in the GS. In recent years, there has been an expansion of higher education institutes in these nations without proper planning, infrastructure, human resources, and checks on quality [8]–[10], [12]–[14], [55], [56]. This has led to unemployed youths with engineering degrees. The reports [7], [57]indicate that less than 50% percent of the graduating students are employable [9], [10], [58]. To tackle these issues, having the right curriculum with properly trained faculty is essential. There is a need to bring innovative ways to retain students' engagement in class. Building strong research capacity is another aspect that will help improve teaching with a strong emphasis on quality. While these challenges are easy to mention on paper, the problems rooted are much more profound, often at the core organizational level. The policies with which these institutes work often lead to poor outputs. Robust hierarchical systems, lack of creative ability to implement new technologies, and poor remunerations or salaries often lead to demotivated employees. Such issues are often undocumented. The effects of these issues are multifaceted and will require close inquiry. To add to these issues, if the Metaverse-based digital transformation is pushed, it will have further consequences. Since digital transformation affects jobs, stakeholders, employees, and the overall dynamics of the workplace, the administration of these higher engineering institutes needs to be more open, experimental, and willing to embrace failures [1]. It will require a thoughtful design process and implementation rooted in social, and organizational issues and quality of work life. As identified from the literature survey, Figure 4 presents some socio-technical challenges to implementing Metaverse-based digital transformation in the GS context.

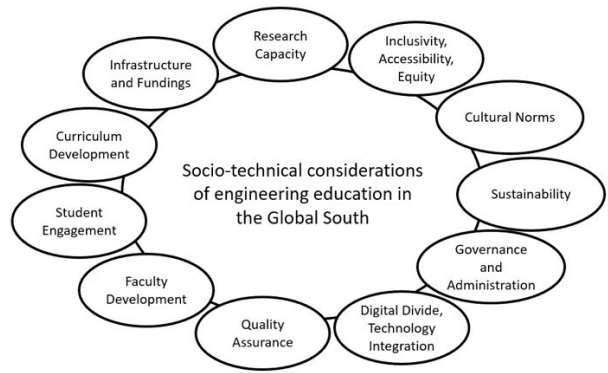

Figure 4: Socio-technical challenges must be addressed for implementing Metaverse-based digital transformation in the GS context.

The literature survey also indicates several HCI concerns that deal with the user experience design, cross-cultural usability, and interaction design aspects of the Metaverse. The Metaverse must account for local factors while designing virtual environments, digital avatars, and user interfaces for GS's diverse and culturally vibrant nations. This entails considering traditional elements in virtual environments, capturing expressions, communications, attire, complexion, and other cultural cues. Further, it is crucial to understand that the Metaverse has different layers. Before the rise in XR technologies, the Metaverse was primarily understood in terms of 2D or 3D virtual worlds. Several research studies [59]–[66] have been conducted to provide a strong backbone for understanding social and cultural norms of virtual worlds and digital avatars. It is also to be noted that these researchers have potentially indicated the technical challenges of implementing XR-based Metaverse that primarily involve issues with scaling and usability, such as motion sickness and nausea [3], [67]. XR is still evolving, and the human factor concerns are still far from being addressed. Implementing a head-mounted display (HMD), such as Hololense, HTC Vive, and Meta Quest, is also challenging across masses due to various factors involving cost, expertise, and infrastructure. A study on the Metaverse workspace [67] also indicates that people are not willing to wear the headset for 9 a.m. to 6 p.m. jobs continuously all the time. It also suggests that users become uncomfortable wearing these headsets and vomit, thus taking these HMDs off very soon. Hence, it would be unwise to view the Metaverse as a plug-and-play escape to the virtual world full of magic and opportunities just by putting on a headset.

In that sense, virtual worlds, such as Second Life, Roblox, etc., provide users with better scalability, interoperability, and control. While not fully immersive, these environments provide users with a certain sense of presence [68]–[70]. If implemented at a broader scale, the UNESCO ICT Framework [71] offers suitable directions for teacher training and their skill building towards using ICT. Such guidelines can also be drafted for the Metaverse – that is informed through contextual inquiry, participatory Design, and keeping the requirements of the GS. Figure 5 presents a list of Socio-technical and HCI Design considerations based on the Design for the GS. While the list is not exhaustive, it offers an initial insight collated from synthesizing the published literature in the Metaverse, Education, and the GS field. ChatGPT-4 was utilized for the tabulation of synthesized insights.

| Challenges | Descriptions | Design Considerations | ST Design Considerations | HCI Design Considerations |
|---|---|---|---|---|
| Infrastructure | Limited access to state-of-the-art equipment, technology, and facilities. | Design should prioritize the use of local materials and resources to create affordable, sustainable infrastructure. Solutions should be culturally and contextually appropriate. | Develop low-cost, high-impact technological solutions adapted to local conditions. Promote resource sharing between institutions. | Design of virtual environments and avatars should consider local aesthetic preferences, cultural symbols, and norms to promote usability and acceptance. |
| Digital Divide | Unequal access to digital resources, including internet connectivity and digital devices. | Design initiatives should consider the local availability and cost of digital resources. Solutions such as offline digital content, low-data usage platforms, or local network systems might be more suitable. | Deploy low-cost, efficient connectivity solutions. Implement programs to distribute affordable devices to students and staff. | User interfaces should be designed with simple, intuitive controls to minimize the learning curve for users with limited digital literacy. Also, consider introducing avatar-guided tutorials for first-time users. |
| Curriculum Development | Outdated or irrelevant curriculum that doesn't meet current industry standards or local needs. | Curriculum should be co-designed with local stakeholders including students, faculty, and industry representatives to ensure relevance to local needs and realities. | Incorporate collaborative, participatory design processes with local industries, alumni, and other stakeholders to update and contextualize curriculum. | Virtual learning environments should be co-designed with local stakeholders, incorporating local contexts and cultural elements to enhance relevance and engagement. |
| Faculty Development | Lack of opportunities for faculty development and training, particularly in emerging technologies. | Training programs should be designed considering the local context, resource limitations, and specific needs of faculty in the Global South. Local experts can be leveraged for delivering training. | Create digital platforms for continuous professional development, including MOOCs, webinars, and online workshops. Foster partnerships with institutions in the Global North for faculty exchange programs. | Training programs should include modules on cross-cultural communication, digital etiquette, and avatar-based teaching methods. |
| Student Engagement | Difficulty in engaging students due to large class sizes, traditional teaching methods, and lack of resources. | Learning experiences should be designed with a deep understanding of the socio-cultural context of students. Local languages, customs, and societal norms should be incorporated in the design of engagement strategies. | Implement blended learning models that combine face-to-face and online learning. Use interactive digital tools to facilitate student participation and feedback. | Avatar-based interaction can enhance student engagement by providing a more immersive, interactive learning experience. Culturally relevant avatars can increase identification and engagement. |
| Research Capacity | Limited capacity for high-quality research due to lack of funding, resources, and training. | Research projects should aim to address local challenges and needs. Collaboration with local industries and communities can enhance the relevance and impact of research. | Establish research partnerships with industries, NGOs, and international institutions. Use digital tools to facilitate collaboration, data collection, and data analysis. | Research on cross-cultural usability and human-avatar interaction can provide valuable insights for improving the design of virtual learning environments. |
| Quality Assurance | Challenges in maintaining and improving quality due to lack of resources and systemic issues. | Quality standards and evaluation criteria should be contextually appropriate and co-developed with local stakeholders. This can ensure a more holistic and inclusive approach to quality assurance. | Develop comprehensive, context-specific quality assurance frameworks that consider both technical and social factors. Use digital analytics to monitor and improve quality. | Quality standards should include criteria for cross-cultural usability and effective human-avatar interaction. User feedback and user testing should be integral parts of the design and quality assurance process. |
| Sustainability | Difficulty in ensuring the economic, environmental, and social sustainability of institutions. | Sustainable design principles such as the use of local and renewable resources, low-energy technologies, and socially responsible practices should guide the design of all aspects of institutional operation. | Implement sustainable practices in all aspects of institutional operation, from energy use to curriculum design. Use digital tools to monitor and manage sustainability. | The sustainability of virtual learning environments should be evaluated not only in terms of environmental impact but also in terms of social and cultural sustainability. |
| Inclusivity | Inequalities in access and participation due to socio-economic, gender, or other barriers. | Design should aim to address local inequalities and barriers to access. This might involve designing for diverse languages, cultural practices, economic situations, and physical abilities. | Develop inclusive policies and practices, from scholarship programs to inclusive pedagogy. Use digital tools to reach and support marginalized students. | Design of virtual environments and avatars should consider the diversity of users in terms of culture, language, gender, physical abilities, etc. to promote inclusivity and accessibility. |
| Accessibility | Barriers that prevent people with disabilities from participating fully in educational activities. | Implement universal design principles in the development of educational resources and environments. Use assistive technologies to enhance access for students with disabilities. | Ensuring the system is usable and beneficial for all users | Intuitive interfaces, avatar-guided tutorials |
| Cross-Cultural Interactions | The potential for misunderstandings and conflicts due to cultural differences among students and staff. | Integrate intercultural communication training into curriculum and staff development. Use digital tools to facilitate cross-cultural collaboration and | Managing the social dynamics within the Metaverse | Culturally appropriate avatar designs, respectful interaction norms |

Figure 5: Design considerations for educational Metaverse in the Global South

## 5  LIMITATIONS OF THE STUDY

This article presents a conceptual insight into the educational Metaverse from the GS perspective. ChatGPT 4.0 was utilized, which prompted queries based on the authors' understanding and insights gained from the literature review. ChatGPT 4.0 was only trained till September 2021. Hence, the insights generated may not be up to date. However, the results synthesized were insightful and have been integrated wherever possible. The study is a first step towards understanding the Design of the educational Metaverse for the GS and primarily relies on published literature. Since the literature in this area is currently sparse, a rapid scoping review technique was adopted. Further empirical studies will be required to deepen the understanding from users' perspectives.